\documentclass[
twocolumn,aps,prx,
 amsmath,amssymb,
floatfix,superscriptaddress
]{revtex4-2}

\usepackage{graphicx}
\usepackage{epstopdf}
\usepackage{physics}
\usepackage{nicematrix}
\usepackage{bbold}
\usepackage{bm}
\usepackage{appendix}
\usepackage{amsmath}
\usepackage[colorlinks = true,
            linkcolor = blue,
            urlcolor  = blue,
            citecolor = blue,
            anchorcolor = blue]{hyperref}

\begin{document}
\title{Superextensive charging speeds in a correlated quantum charger}
\author{Harald Schmid}\email{h.schmid@tum.de}
\affiliation{\mbox{Technical University of Munich, TUM School of Natural Sciences, Physics Department, 85748 Garching, Germany}}
\affiliation{Munich Center for Quantum Science and Technology (MCQST), Schellingstr.\ 4, 80799 M{\"u}nchen, Germany}
\affiliation{\mbox{Dahlem Center for Complex Quantum Systems and Fachbereich Physik, Freie Universit\"at Berlin, 14195 Berlin, Germany}}
\author{Felix von Oppen}\email{vonoppen@physik.fu-berlin.de}
\affiliation{\mbox{Dahlem Center for Complex Quantum Systems, Fachbereich Physik, and Halle-Berlin-Regensburg}\\ Cluster of Excellence CCE, Freie Universit\"at Berlin, 14195 Berlin, Germany}
\author{Gil
Refael}\email{refael@caltech.edu}
\affiliation{\mbox{Institute of Quantum Information and Matter and Department of Physics,} \mbox{ California Institute of Technology, Pasadena,
CA 91125, USA}}

\author{Yang Peng}\email{yang.peng@csun.edu}
\affiliation{\mbox{Department of Physics and Astronomy, California State University, Northridge, CA 91330, USA}}
\affiliation{\mbox{Institute of Quantum Information and Matter and Department of Physics,} \mbox{ California Institute of Technology, Pasadena,
CA 91125, USA}}

\begin{abstract}
We define a quantum charger as an interacting quantum system that transfers energy between two drives. The key figure of merit characterizing a charger is its charging power. Remarkably, the presence of long-range interactions within the charger can induce a collective steady-state charging mode that depends superlinearly on the size of the charger, exceeding the performance of noninteracting, parallel units. Using the driven Lipkin–Meshkov–Glick model and power-law interacting spin chains, we show that this effect persists up to a critical system size set by the breakdown of the high-frequency regime. We discuss optimal work output as well as experimentally accessible initial states. The superlinear charging effect can be probed in trapped-ion experiments, and positions interacting Floquet systems as promising platforms for enhanced energy conversion.
\end{abstract}
\maketitle
\date{\today}

\section{Introduction}

A critical engineering objective for developing quantum technology is controlling the energy flow in quantum devices \cite{Lauk2020,Lambert2020}. Given that quantum platforms are based on diverse physical implementations \cite{Loss1998,Burkard2023,Koch2007,Kjaergaard2020,Cirac1995,Bruzewicz2019}, the coherent transfer of energy is essential for inter-platform crosstalk \cite{Tanzilli2005,Kimble2008}. 
 
Floquet engineering provides a practical tool to bridge this communication gap \cite{Bukov2015,Oka2019}. Under strong periodic drives, quantum systems can act as pumps that transfer a quantized amount of energy per period (in units of the driving frequencies) \cite{Martin2017,Peng2018,Nathan2019,Kolodrubetz2018,
Nathan2021,Peng2025}. In the example of a single qubit with two incommensurately oscillating fields, the work quantization can be understood via mapping to a topological insulator on a two-dimensional frequency lattice. Viewing one of the drives as an energy source, and the second drive as a battery (realized as an oscillator storing energy), the resulting system realizes a \textit{quantum charger} which operates in a Floquet steady state. Recent theoretical work on energy conversion with periodic drives has focused on questions of nonadiabaticity \cite{Long2021,Qi2021,Psaroudaki2021,Psaroudaki2023}, and proposed implementations in Weyl semimetals \cite{Nathan2022} as well as  rhombohedral multilayer graphene \cite{Lantagne2024}. An experiment on superconducting qubits was reported in Ref.\ \cite{Malz2021}. However, these studies were restricted to noninteracting systems.

\begin{figure}[b!]
    \centering
    \includegraphics[width=\linewidth]{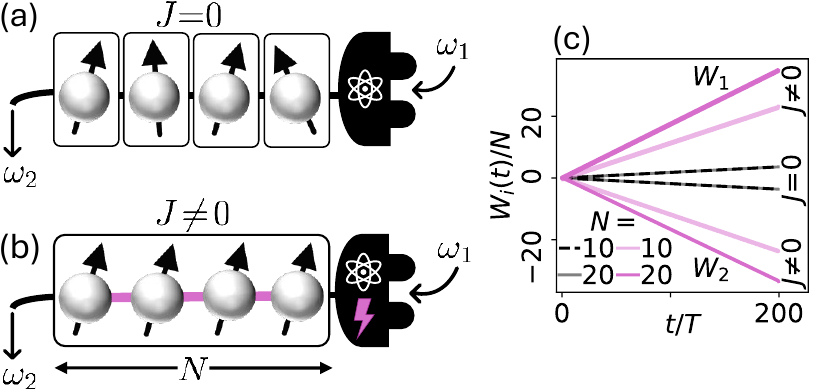}
    \caption{Energy flow between drives with frequencies $\omega_1$ and $\omega_2$ through a many-body quantum charger. Arrows indicate the direction of the energy flow (sign of the work). (a) Parallel, noninteracting  charger. (b) Collective, interacting charger. (c) Work per spin in a Floquet steady state. Long-range interactions enhance the charger's performance superlinearly in the system size $N$. }
    \label{fig:charger}
\end{figure}

Here, we generalize energy conversion with periodic drives to interacting many-body quantum systems. Our focus is on nonequilibrium steady states, so that
we work within a Floquet description, which remains applicable in the interacting case. The system then realizes a \textit{many-body quantum charger}, capable of transferring a large number of quanta (``photons") between the drives during each period. 

Our central objective is to investigate whether the presence of interactions can parametrically enhance the performance of the charger. We  demonstrate that the transferred energy $W$ can scale as
\begin{align}
    W \propto N^{1+\delta}
\end{align}
over a broad range of system sizes $N$, with $\delta>0$, when interactions are present. This is schematically illustrated in Fig.~\ref{fig:charger}, where we compare the pumping of independent, noninteracting chargers [Fig.~\ref{fig:charger}(a)] to a collective charger in the presence of interactions [Fig.~\ref{fig:charger}(b)]. 

Our notion of a quantum charger is related to quantum batteries, systems designed for energy storage and release \cite{Alicki2013,Hovhannisyan2013, Binder2015b,Ferraro2018,Andolina2018,Andolina2019,Farre2020,Gyhm2022,Rossini2020,Campaioli2024,Gyhm2024,Romero2025}. Previous work has shown that many-body batteries can harness collective effects \cite{Ferraro2018,Rossini2020,Campaioli2024}. Here we demonstrate that chargers exhibit superextensive charging rates due to many-body interactions, with the work per Floquet period $W$ providing a natural measure of the effective charging speed. We emphasize, however, that the operational principles of the two devices differ in important ways: Batteries are transient systems that charge over a finite time. In contrast, the charger continuously pumps energy between the drives in a nonequilibrium Floquet steady state. As a consequence of its transient nature, batteries require optimized charging protocols \cite{Mazzoncini2023,Rodriguez2024}. In contrast, the steady-state charger functions without the need for a protocol. We note that our proposal also relates to fractional Thouless pumps \cite{Grusdt2014,Taddia2017,Walter2023,ArguelloLuengo2024,Juergensen2025,Viehbahn2024} and interacting quantum motors \cite{Bustos2013,Arrachea2015,Bruch2018,Arrachea2023,Calvo2017}.

In our analysis, we model  a quantum charger as a long-range interacting \cite{Campa2009,Defenu2023}  spin chain subject to two classically oscillating fields. For all-to-all interactions (Sec.\ \ref{sec:Superlinear charging}), the model realizes a driven Lipkin-Meshov-Glick (LMG) model \cite{Lipkin1965}, whose nonequilibrium dynamics has attracted significant attention  \cite{Das2006,Caneva2008,Jörg2010,Sciolla2011,Campbell2016,Russomanno2017,Zhou2017,Pappalardi2018,Ferreira2019,Pizzi2021,Block2022,Afrasiar2023,Bento2024,Zhou2024,Santini2025}.
We demonstrate that  in its optimal operating modes, the charger converts energy superlinearly between the drives, both over a period and in the steady state [Fig.~\ref{fig:charger}(c)]. For all-to-all coupling, the work scales quadratically up to a characteristic system size $N^\ast$, after which the scaling crosses over to linear. This crossover size can be estimated by comparing the interaction scale with the driving frequency, and  
marks the transition between the high-frequency and the low-frequency regime.  We show that a high-frequency expansion of the Floquet operator accurately captures the work in the superlinear scaling regime  (Sec.\ \ref{sec:High-frequency, adiabatic limit}). Superlinear charging is most efficient for comparable drive frequencies and decays exponentially with the driving period when they differ. We establish the effect for a wide range of realistic long-range interactions and provide corresponding scaling estimates. In Sec.\ \ref{sec:Stabilization of superlinear pumping}, we identify accessible initial states to observe the scaling experimentally. Section \ref{sec:conclusion} contains concluding remarks.

\section{Superlinear charging}
\label{sec:Superlinear charging}
We model a quantum many-body charger by the time-dependent Hamiltonian
\begin{align}
    H(t)=H_0 + V_1(\omega_1t + \phi_1)+V_2(\omega_2t+\phi_2).
    \label{eq:Hamiltonian}
\end{align}
The charger transfers energy between the two drives $V_{i}$ with $i=1,2$ through the static part $H_0$. Each drive oscillates periodically in time with frequency $\omega_{i}$ with associated individual period $T_i=2\pi/\omega_i$ and phase $\phi_{i}$. We focus on periodic driving characterized by a commensurate ratio $\omega_1/\omega_2=p/q$, where $p$ and $q$ are coprime integers. This choice fixes the overall period as $T=pT_1=qT_2$, with fundamental frequency $\omega=2\pi/T$. We treat the dynamics in a Floquet picture where the quasi-stationary Floquet states $\ket{\psi_\alpha}$ are defined through the eigenvalue equation $U(T)\ket{\psi_\alpha}=e^{-i\epsilon_\alpha T} \ket{\psi_\alpha}$ of the time-evolution operator $U(t)=\mathcal{T}e^{-i\int^t_0 dt^\prime H}$ for a single Floquet period $T$. The quasi-energies can be restricted to the first Floquet zone $-\omega/2 < \epsilon_\alpha\leq \omega/2$.  

The integrated pumping power (work) from each drive is given by
\begin{align}
W_{i}(t)&=\int^t_0 \dd{t'}\ev*{\dv{V_i(t')}{t'}} {\psi(t')}
\notag
\\
&=i\omega_{i}\ev*{U^{\dagger}(t)\pdv{U(t)}{\phi_{i}}}
{\psi(0)}. 
\label{eq:work operator 2}
\end{align} 
Here, the state $\ket{\psi(t)}=U(t)\ket{\psi(0)}$ of the system is time-evolved from the initial state $\ket{\psi(0)}$. The total work transferred over a single Floquet period, $W_i(T)$, thus provides a natural measure of the charging rate.

From the second line in Eq.\ \eqref{eq:work operator 2}, we identify the (hermitian) single-period work operator \cite{Psaroudaki2023}
\begin{align}
\mathcal{W}_{i}(T)=i\omega_{i}U^{\dagger}(T)\pdv{U(T)}{\phi_{i}},
\label{eq:work operator}
\end{align}
whose expectation value for a given initial state determines the work output. It is useful to define the work spectrum $\mathcal{W}_{i}\ket{w_{i,n}} = w_{i,n}\ket{w_{i,n}}$. The maximum possible work over one period is obtained by initializing the system in the optimal work state $\ket{w_{i,n}}$ corresponding to the largest work eigenvalue, $\mathrm{max}_{n}w_{i,n}\equiv||\mathcal{W}_{i}||$, where $\|\cdot\|$ is the operator norm.

However, work states do not return to themselves after one driving period. Hence, their work output cannot be generated in the steady state. To get a quasi-stationary work output, we also consider initialization in a Floquet state. The work output per cycle for Floquet states is
\begin{align}
    W_{i,\alpha}(T)=\omega_i T\pdv{\epsilon_\alpha}{\phi_i}
    =\ev{\mathcal{W}_i}{\psi_\alpha}.
\end{align}
Because steady states produce no net work on average, the energy from the first drive is fully transferred to the second, $W_{1,\alpha}(T)=-W_{2,\alpha}(T)$.

\begin{figure}[t!]
    \centering
    \includegraphics[width=\linewidth]{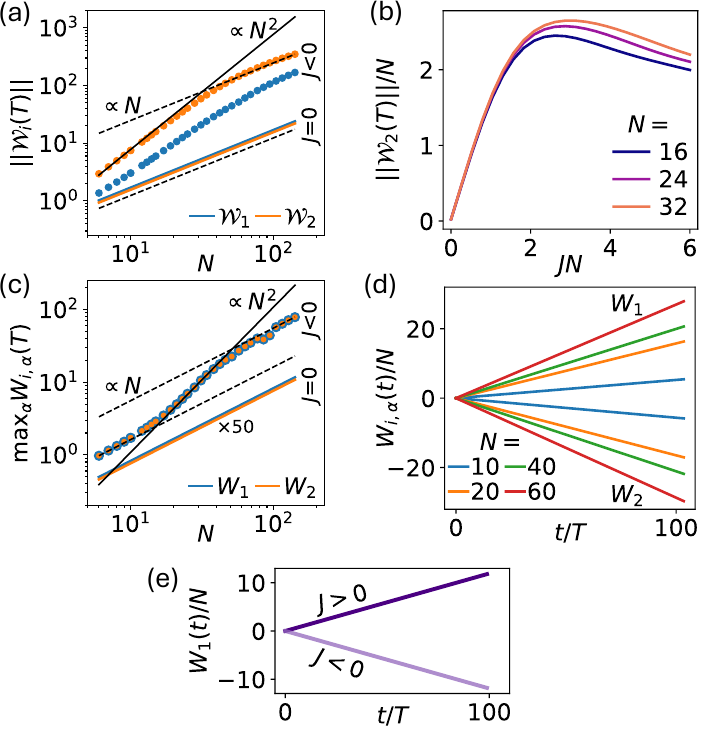}
    \caption{Superlinear work scaling with $N$ of an interacting quantum charger. (a) Maximum work over a period scales quadratically in $N$ in the presence of interactions (dots), crossing over to linear scaling at large $N$. The noninteracting charger (solid) always scales linearly. (b) Same as (a) depending on the rescaled couplings. (c) Work of optimal Floquet state also scales superlinearly for intermediate $N$. (d) Superlinear steady-state energy flow via dynamics of Floquet states (evaluated at multiples of the period). (e) The sign of the interaction controls the direction of energy flow for Floquet states. Parameters: $\omega_1=2\pi$, $\omega_2=\pi$, $\phi_1=0$, $\phi_2=\pi/4$,
$\mathbf{h}_1=0.1\hat{\mathbf{y}}+\hat{\mathbf{z}}$ 
$\mathbf{h}_2=\hat{\mathbf{x}}+0.1\hat{\mathbf{y}}$, (a),(c) $J=-0.05$, $\mathbf{h}_0=0.01\hat{\mathbf{x}}$ (d),(e) $J=-0.02$, $\mathbf{h}_0=0.25\hat{\mathbf{x}}$, $N=10$.}
    \label{figure1}
\end{figure}

As a concrete example system, we consider the spin chain
\begin{align}
H_{0}&=\sum_{i<j}^{N}J_{ij}S_{z}^{(i)}S_{z}^{(j)}
+\sum_{j=1}^{N}\mathbf{h}_{0}\cdot \bm{S}^{(j)},
\notag
\\
V_{i}&=\cos(\omega_i t+\phi_i)\sum_{j=1}^{N}\mathbf{h}_{i}\cdot \bm{S}^{(j)},
\label{eq:spin chain}
\end{align}
with $N$ spin-$\frac{1}{2}$ operators $S_{\alpha=x,y,z}^{(j)}$. The spins couple via static Ising couplings $J_{ij}$ and are subject to a magnetic field with a static $\mathbf{h}_0$ and two driven components $\mathbf{h}_{1,2}$. The work can be expressed by integrating the spin dynamics modulated by the driving field, 
\begin{align}
W_{i}=-\omega_i\sum^N_{j=1}\int^t_0 \dd{t'}\mathbf{h}_{i}\cdot\ev*{\bm{S}^{(j)}}{\psi(t')}\sin(\omega_it'+\phi_i).
 \end{align}
 
 We first focus on all-to-all couplings $J_{ij}=J$. In this case, it is useful to define collective spin operators, $\bm{S} = \sum_{i}\bm{S}^{(i)}$. The spin chain realizes a driven Lipkin-Meshov-Glick (LMG) model
 \begin{align}
H(t)&=\frac{J}{2}S_z^2+\mathbf{h}_0 \cdot \mathbf{S}+\sum^2_{i=1}\cos(\omega_it+\phi_i)\mathbf{h}_i \cdot \mathbf{S}.
\end{align}
with conserved total spin $\mathbf{S}^2=S_x^2+S_y^2+S_z^2$. We specify to the maximum total spin $S=N/2$ and calculate observables in the  $\ket{S_z}$ basis with magnetic quantum number $S_z=-N/2,\dots,N/2$. The LMG model has been experimentally realized with Josephson junctions of Bose-Einstein condensates \cite{Albiez2005} as well as  superconducting qubits \cite{Xu2020}. 

Our main goal is to characterize the scaling of the work with the number of spins $N$. To calculate the work operator and its spectrum, we numerically time-slice the Floquet operator and discretize the derivatives $\partial_{\phi_i}$. Figure \ref{figure1} shows the work performed by the individual drives for the all-to-all interacting charger with driving ratio $p/q=2$ (see legend for other parameters). In the presence of interactions, we observe that  for small system sizes, the maximum work $||\mathcal{W}_i(T)||$ associated with the optimal work state shows a quadratic scaling  $\propto N^2$ [Fig.~\ref{figure1}(a)]. In this regime, the charger effectively transfers the interaction energy $I=\sum_{i,j}J_{ij}=JN^2$ from one drive to the other. For larger system sizes, we observe a crossover to linear scaling, $\propto N$. In contrast in the noninteracting case, the work always depends linearly on $N$. This is consistent with the expectation that each spin evolves independently from the others, so that the work of the drives is just the sum of contributions from the individual spins. 

We stress that the nonlinear scaling with $N$ is nontrivial in the all-to-all interacting model Eq.\ \eqref{eq:spin chain} since the drives couple linearly in $N$ to the charger. Note that the data in Fig.~\ref{figure1}(a) are for small ferromagnetic couplings $J=-0.05$, but the effect is insensitive to the sign of $J$ for work states. We further comment that the difference between $||\mathcal{W}_1(T)||$ and $||\mathcal{W}_2(T)||$ for $J<0$ does not indicate any heating. Since the work states used to define the maxima are not quasi-stationary, there is no requirement that the total work over a single cycle sums to zero. 
\begin{figure}[t!]
    \centering
    \includegraphics[width=0.45\textwidth]{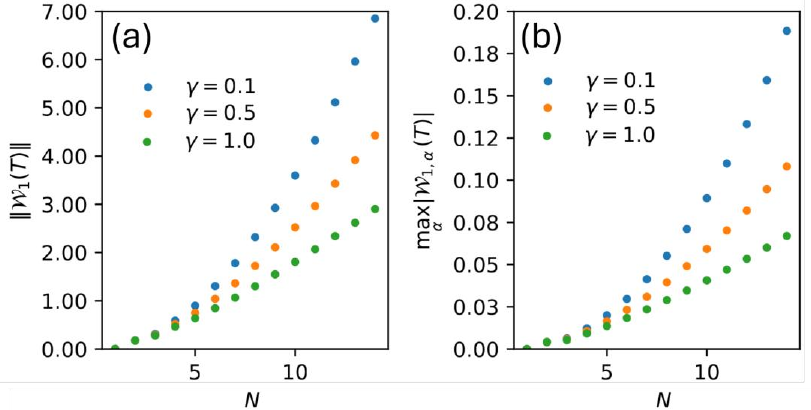}
    \caption{(a) Maximum work, and (b) work of optimal Floquet state, for drive 1 over a period as a function of system size for various power-law interactions with range $\gamma$.  Parameters: $J=1$, $T=0.1$, $\mathbf{h}_0=0$, $\mathbf{h}_1=0.86\mathbf{\hat{x}}+0.25\mathbf{\hat{y}}+0.41\mathbf{\hat{z}}$, $\mathbf{h}_2=0.18\mathbf{\hat{x}}+0.67\mathbf{\hat{y}}+0.72\mathbf{\hat{z}}$, $\phi_1=0$, $\phi_2=\pi/2$.}
    \label{fig:long_range}
\end{figure}

Figure \ref{figure1}(b) plots the work of the second drive versus the rescaled coupling strength $JN$. For small $JN$, we identify a linear scaling of  the work per spin, $||W_2(T)/N||$, with $N$, with data from different system sizes collapsing onto a single curve. For large $JN$, $||W_2(T)/N||$ saturates to a constant, consistent with our findings in Fig.~\ref{figure1}(a). This transition occurs when the available interaction energy becomes comparable to the energy scale set by the driving period, $JN^2\sim 2\pi/T$. 

\begin{figure*}[t!]
    \centering
    \includegraphics[width=\textwidth]{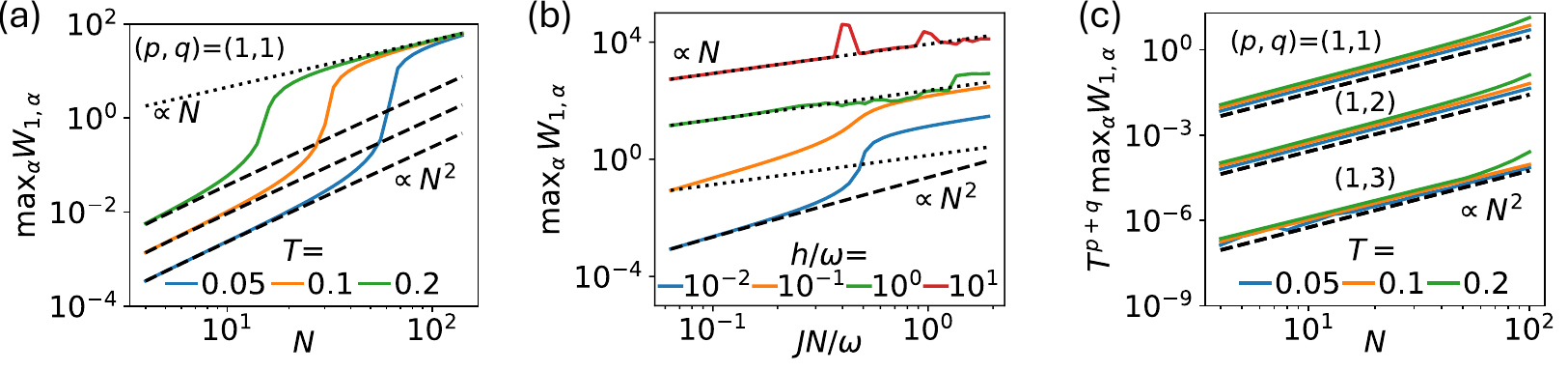}
    \caption{Maximum work of Floquet states in the high-frequency regime. (a) $\omega_1=\omega_2$. Analytical expression in Eq.\ \eqref{eq:high freq analytical} (dashed) captures data for small $JN\ll \omega$. (b) Same as in (a) but for different amplitudes $h/\omega$, see legend. (c) Various frequency ratios $p/q=\omega_1/\omega_2$ as indicated. We collapse the data according to Eq.\ \eqref{eq:work freq scaling}. Parameters: (a) $J=1$, (b) $J=1$, $T=0.1$, (c) $J=0.1$. Other parameters as in Fig.~\ref{fig:long_range}.}
    \label{fig:high_freq}
\end{figure*}

We now analyze Floquet states to investigate whether superlinear pumping is possible in the steady state. We identify the optimal Floquet state by maximizing the expectation value of the work operator. Figure \ref{figure1}(d) shows its work dynamics for the first drive. Our results confirm that, even in the steady state, the work per period can depend superlinearly on $N$. As for work states, Floquet states undergo a transition from quadratic to linear $N$-scaling [Fig.~\ref{figure1}(c)], with similar crossover system size. We note, however, that work states can pump a substantially larger absolute amount of work per period (about an order of magnitude for the parameters in Fig.~\ref{figure1}). Interestingly, the direction of energy flow for Floquet states can be controlled via the sign of the interaction $J$ in the superlinear regime [Fig.~\ref{figure1}(e)]. We have checked this holds independent of the specific $p,q$, and follows from the linear dependence $\propto JN^2$ of the work on the interactions  in the superlinear regime.

The superlinear pumping effect holds also for finite-range interactions. We consider the more realistic power-law interactions $J_{ij} = J/|i-j|^{\gamma}$, which  can be realized in trapped-ion experiments \cite{Britton2012,Islam2013,Zhang2017} with tunable range $0\leq \gamma\leq 3$. 
The results of our numerical calculation are shown in Fig.~\ref{fig:long_range}. The maximum work and work of optimal Floquet states show a superlinear scaling effect for $\gamma<1$. To get the $N$-scaling, we estimate the available interaction energy, which is
\begin{align}
    W\propto N\sum^N_{j>0}\frac{1}{|j|^{\gamma}} \propto\begin{cases}
        N^{2-\gamma} \qquad &\gamma<1,
        \\
        N\ln N  \qquad &\gamma=1,
        \\
        N \qquad &\gamma>1.
    \end{cases}
\end{align}
In App.~\ref{app:fintie_interaction}, we plot the results in log scale to compare with the scaling estimate, which is consistent with our data in Fig.~\ref{fig:long_range}. For the rest of the paper, we focus on all-to-all couplings. 

The transition can be understood from the properties of the Floquet many-body spectrum. Since the work performed by a Floquet state is determined by its phase derivative, the scaling of the optimal work with $N$ is governed by the scaling of the many-body bandwidth, provided that quasienergies depend smoothly on the phase. For $JN,h\ll\omega$, the Floquet spectrum covers only a fraction of the Floquet Brillouin zone. In this regime, the bandwidth is not constrained by the periodicity of the quasienergies and can therefore grow superlinearly with system size. This simple picture leads to the superlinear scaling of the work. In contrast, when $JN^2\ll \omega$ or at $h\gtrsim\omega$, the many-body quasienergy spectrum extends across the entire Floquet Brillouin zone, $-\pi/T\lesssim \epsilon_\alpha \lesssim \pi/T$. Since quasienergies are defined modulo $\omega$, the bandwidth cannot increase indefinitely. In other words, once the interaction energy exceeds the size of the Floquet Brillouin zone, the system can no longer distinguish energies separated by integer multiples of $\omega$. Thus, the growth of the work slows down and eventually crosses over to linear scaling with $N$.

\section{High- and low-frequency limits}
\label{sec:High-frequency, adiabatic limit}
\subsection{High-frequency limit}
The superlinear scaling of the work at small system sizes can be understood in the high-frequency limit $JN,h \ll \omega$. Figure \ref{fig:high_freq}(a) displays the maximum work of Floquet states when the two driving frequencies are equal ($p=q=1$). In this case, the work scales quadratically, $W\propto N^2$, provided that $JN \ll \omega$. In contrast for $JN \gg \omega$, the scaling becomes linear $W\propto N$, indicating a crossover at $JN \sim \omega$. Interestingly, and in contrast to the intermediate-frequency regime shown in Fig.~\ref{figure1}(c), the work can exceed quadratic scaling in this crossover regime. The system exits the superlinear regime also for large amplitudes $h\gtrsim \omega $ [Fig.~\ref{fig:high_freq}(b)], reinforcing that the high-frequency limit must be satisfied.

A systematic comparison across various frequency ratios $p/q$ is shown in Fig.~\ref{fig:high_freq}(c). We find that the data collapses well for various $p/q$ when rescaled according to
\begin{align}
    W_{i,\alpha}(T)\propto \frac{N^2}{T^{p+q}}.
\label{eq:work freq scaling}
\end{align}
The dependence on the period $T$ indicates that efficient nonlinear pumping is achievable only for small $p$ and $q$, but drops exponentially with increasing $p+q$.

\begin{figure}[t!]
    \centering
    \includegraphics[width=\linewidth]{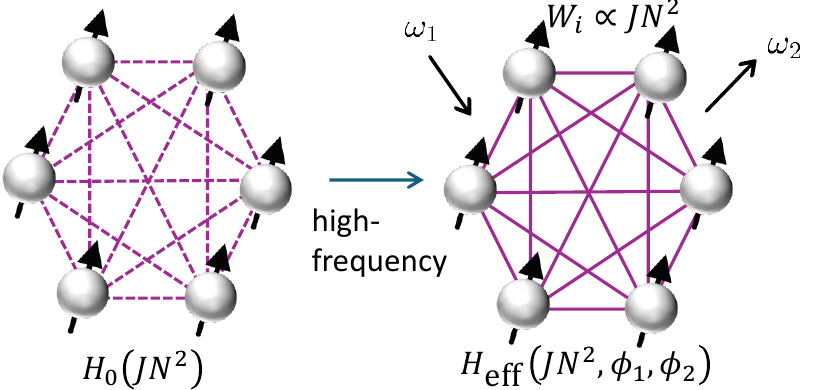}
    \caption{Static all-to-all spin model (left) vs. dynamic all-to-all charger (right). While the bare interaction $H_0$  (dashed lines) is all-to-all and thus intrinsically nonlinear, they do not directly contribute to the charging process and thus do not guarantee superlinear energy transfer. In contrast, the transferred energy in the driven system is governed by the effective high-frequency Hamiltonian $H_\mathrm{eff}$, which depends on the drive phases $\phi_i$ and enables the nonlinear charging effect mediated via effective interactions (solid lines).}
    \label{fig:all-to-all hf Hamiltonian}
\end{figure}

While the static all-to-all interaction $H_0$ is already nonlinear in system size, it does not by itself guarantee nonlinear energy transfer. Instead, the relevant contribution arises from an effective high-frequency Hamiltonian, illustrated schematically in Fig.~\ref{fig:all-to-all hf Hamiltonian} and calculated in the following.
We perform a van Vleck expansion in the high-frequency limit  \cite{Eckardt2015,Ikeda2021}, derived from writing the Floquet operator as $U(T)=e^{-iK}e^{-iH_\mathrm{eff} T }e^{iK}$ with effective Hamiltonian  $H_\mathrm{eff}$  and kick operator $K$. For Floquet states, the work is given by the $\partial_{\phi_i}$-derivative of the effective Hamiltonian 
\begin{align}
W_{i,\alpha}(T)=
\omega_i T\ev*{\pdv{H_\mathrm{eff}}{\phi_i}}{u_\alpha}.
\label{eq:work Heff} 
\end{align}
Here, $\ket{u_\alpha}=e^{iK}\ket{\psi_\alpha}$ are eigenstates of $H_\mathrm{eff}$.
Up to second order, the high-frequency expansion of $H_\mathrm{eff}$ in powers of $1/\omega$ yields  \cite{Eckardt2015,Ikeda2021}
\begin{align}
    &H^{(0)}_\mathrm{eff}=H_0,  \quad\quad
     H^{(1)}_\mathrm{eff}= \sum_{n\neq 0} \frac{[H_{-n},H_n]}{2n \omega},
     \notag
     \\
     &H^{(2)}_\mathrm{eff}=\sum_{n\neq 0}\frac{[[H_{-n},H_0],H_n]}{2n^2 \omega^2}
    +\sum_{ n,m\neq 0 }\frac{[[H_{-n},H_{n-m}],H_n]}{3nm \omega^2}, 
    \label{eq:Heff}
\end{align}
The successive commutators contain the Fourier components in $H(t)=\sum^\infty_{n=-\infty}e^{-in\omega t}H_n$. Due to the all-to-all interactions, the bandwidth of the effective Hamiltonian scales superlinearly with system size, $D_{\mathrm{eff}}\propto N^2$. However, the work in Eq.\ \eqref{eq:work Heff} depends on the phase derivative, so a superlinear contribution can only be generated at higher orders in $1/\omega$. Specifically, $H_0$ is superlinear in $N$ but independent of $\phi_i$. In contrast, the drives (associated to $H_n$ with $n\neq 0$) are linear in $N$ and depend on $\phi_i$.  As a result, superlinear behavior can emerge at second order in $1/\omega$, provided that both drives have the same frequency ($p=q=1$), in agreement with Fig.~\ref{fig:high_freq}(a). 

For other ratios $p\neq q$, the superlinearity only appears at higher orders in the effective Hamiltonian, consistent with the data collapse proposed in Eq.\ \eqref{eq:work freq scaling}. Each term in the van Vleck expansion involves nested commutators of Fourier components $H_n$, with the constraint that the sum of all indices vanishes.  Physically, this reflects conservation of the net number of absorbed and emitted “photons” from the drives. A nonlinear term in $H_{\mathrm{eff}}$ requires  mixing $p$ and $q$ harmonics of the two drives, which combine $H_0$ with nontrivial $\phi_i$-dependence that survives the $\partial_{\phi_i}$-derivative. For a ratio $p/q$, the lowest-order process that generates such phase dependence occurs at order $p+q$ in the expansion. We explicitly provide further justification for the scaling in the case $p=1$, $q=2$, see App.\ \ref{app:high freq work}.

We explicitly evaluate the work in the case $\omega=\omega_1=\omega_2$. We find for the effective Hamiltonian  
\begin{equation}
  H^{}_\mathrm{eff} \approx JS_z^2 + \frac{2}{\omega}h^{(1)}_\mathrm{eff} + \frac{J}{\omega^2}h^{(2)}_\mathrm{eff}  
\end{equation}
up to second order in $1/\omega$. The first and second contributions with negative powers of $\omega$ involve
\begin{equation}
   h^{(1)}_\mathrm{eff} =    \sin(\Delta\phi_{12})\mathbf{S}\cdot (\mathbf{h}_1 \times \mathbf{h}_2 )
\end{equation}
and 
\begin{align}
    &h^{(2)}_\mathrm{eff}=\sum^2_{i,j=1}\cos(\Delta \phi_{ij})\bigg[
    -(h_{i,y}h_{j,x}+h_{j,y}h_{i,x})
    \{S_y,S_x\} \notag \\
    &+  
    \bigg(
    2h_{i,y}h_{j,y}
    (S^2_x-S_z^2) 
    +h_{i,z}h_{j,y}\{S_z,S_y\}+(x\leftrightarrow y)\bigg)
    \bigg]
    \label{eq:high freq analytical}
\end{align}
with phase difference $\Delta\phi_{ij}=\phi_i-\phi_j$.
We take the expectation value in Eq.\ \eqref{eq:work Heff} in the unperturbed eigenstates $\ket{u_{\alpha}}\approx \ket{S_z}$. This yields for their work
\begin{align}
&W_{i,S_z}(T)
\approx
\pm \frac{2S_z}{\omega}
(h_{1x}h_{2y}-h_{1y}h_{2x})
\cos\Delta\phi_{12}
\notag\\
&\quad \mp
\frac{2J}{\omega^2}
\Big[
\frac{N}{2}\Big(\frac{N}{2}+1\Big)-3S_z^2
\Big]
(h_{1x}h_{2x}+h_{1y}h_{2y})
\sin\Delta\phi_{12}.
\end{align}
where the positive (negative) sign appears for drive $i=1$ ($i=2$). To suppress the first order, one can choose a phase difference $\Delta \phi_{21}=\pi/2$, or the directions of the fields such that their cross product has no component parallel to the interaction. In that case, the maximum work is achieved for $S_z=N/2$. Our analytical calculations capture the numerical result in Fig.\ \ref{fig:high_freq}(a) in the high-frequency limit (small $N$). We  reproduce the reversal of the energy flow by changing the sign of the interaction $J$ at leading order in $JN$.

\begin{figure*}[t!]
    \centering
    \includegraphics[width=\textwidth]{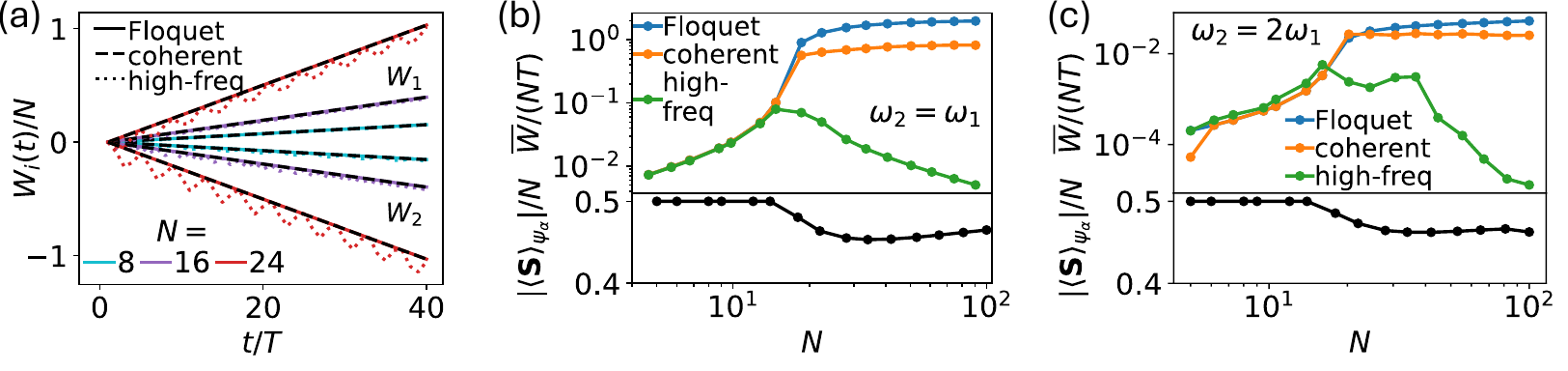}
    \caption{Work for high-frequency and coherent states approximations of Floquet states (experimentally preparable). (a) High-frequency regime ($\omega_1=\omega_2$): Superlinear steady-state pumping is observable both by the high-frequency approximation (dotted) and by the spin-coherent state approximation (dashed) of the optimal Floquet state (solid).
    (b), (c) Average work per cycle and per spin for different frequency ratios $\omega_1/\omega_2=p/q$ as indicated (top panels). Bottom panels: Total polarization  per spin in the Floquet state. Parameters as in Fig.~\ref{fig:long_range}.}
\label{fig:Floquet_vs_coherent_vs_high_freq}
\end{figure*}

\subsection{Low-frequency limit}
In the large $N$ limit, the work always becomes linear in system size, see Fig.\ \ref{figure1}. Here, we show this explicitly in the low-frequency limit $\omega \ll JN,h$, using a semiclassical argument. We also assume $h\ll JN$. We approximate the spin dynamics by replacing collective spin operators by numbers $\mathbf{S}_i\rightarrow \mathbf{s}_i$, leading to the Landau-Lifshitz equation of motion
\begin{align}
    \dot{\mathbf{s}}= \mathbf{h}_\mathrm{eff}(t)\times \mathbf{s}.
    \label{eq:LLG}
\end{align}
The classical spin has length $|\mathbf{s}|=N/2$ and precesses in the effective field 
$\mathbf{h}_\mathrm{eff}(t)=2Js_z \hat{\mathbf{z}}+\mathbf{h}(t)$, which contains the exchange  field alongside the external field $\mathbf{h}(t)=\mathbf{h}_0 +\sum^2_{i=1}\cos(\omega_it+\phi_i)\mathbf{h}_i$. 

The work can then be computed from the spin trajectory  
\begin{align}
    W_i(t)=\omega_i \int\dd{t} 
(\partial_{\phi_i}\mathbf{h}_{\mathrm{eff}})\cdot
\mathbf{s}(t). 
\end{align}
The phase derivative of the effective field contains two contributions, $\partial_{\phi_i}\mathbf{h}_{\mathrm{eff}}=2J(\partial_{\phi_i}s_z)\hat{\mathbf{z}}+\partial_{\phi_i}\mathbf{h}$, but only the exchange-field term depends on $N$. One finds for the nonlinear contribution to the work 
\begin{align}
    W_{i,\mathrm{NL}}(t)=4J\omega_i\int\dd{t}\partial_{\phi_i}s_z^2(t).
\end{align}
For large $N$ and at leading order in $\omega$, one expects that the spin remains nearly locked in the direction of the exchange field. A self-consistent solution of the equations of motion yields [App.\ \ref{app:Low-frequency limit: semiclassical dynamics}]
\begin{align}
\mathbf{s}(t)\simeq \frac{1}{2J}
    \big[
        h_x(t)\mathbf{\hat{x}}
        +
        h_y(t)\mathbf{\hat{y}}
     +
     \left(JN+h_z(t)\right)\mathbf{\hat{z}}\big] +
     \order{\omega}.
\end{align}
The longitudinal component $s_z$ becomes effectively constant in time (at leading order in $N$) and thus independent of $\phi_i$. Hence, the nonlinear contribution $W_{i,\mathrm{NL}}$ vanishes in the low-frequency limit. In App.\ \ref{app:Low-frequency limit: semiclassical dynamics}, we extend the self-consistent solution of the equation of motion to order $\omega$ and show that the nonlinear contribution to the work vanishes even at this order.

\section{Stabilization of superlinear pumping}
\label{sec:Stabilization of superlinear pumping}

\subsection{Superlinear steady state pumping  beyond Floquet states}
\label{sec:high-freq and coherent states}

In general, Floquet states are challenging to prepare experimentally, as they are typically highly fine-tuned. 
It is therefore important to ask whether more easily preparable states can also show superlinear pumping over many cycles. In the following, we consider high-frequency states and spin-coherent states as approximations to the optimal Floquet state. High-frequency eigenstates are eigenstates to the effective high-frequency Hamiltonian. For $\omega_1=\omega_2$ and at lowest order in $\omega$, they are just $\ket{S_z}$-states. Optimal pumping is then achieved for $\ket{S_z=N/2}$. Additionally, the mean-field spin nature of the LMG model motivates us to study spin-coherent states
\begin{align}
    \ket{\phi,\theta}=e^{-i\phi S_z}e^{-i\theta S_y}\big|S_z=\frac{N}{2}\big\rangle
    \label{eq:coherent spin states}
\end{align}
which are parameterized by the polar and azimuthal angles $ \phi$ and $\theta$. We choose the angles by matching the expectation values $\expval{\dots}_\alpha=\expval{\psi_\alpha|\dots|\psi_\alpha}$ to that of the optimal Floquet state, 
$\expval{S_x}_{\alpha}=| \expval{\mathbf{S}}_{\alpha}|\cos(\phi)\sin(\theta)$,
$\expval{S_y}_{\alpha}=| \expval{\mathbf{S}}_{\alpha}|\sin(\phi)\sin(\theta)$ etc.
For the spin-coherent states, the total polarization is maximal $|\expval{\mathbf{S}}|=\sqrt{\expval{S_x}^2+\expval{S_y}^2+\expval{S_z}^2}=N/2$ while for the optimal Floquet states $|\expval{\mathbf{S}}_\alpha|\leq N/2$. Both high-frequency states and spin-coherent state are easy to prepare experimentally.

In Fig.~\ref{fig:Floquet_vs_coherent_vs_high_freq}(a), we compare the work outputs of the high-frequency state and spin-coherent state to that of the Floquet state. For parameters in the high-frequency regime and $\omega_1=\omega_2$, we find for both approximations that there is stable pumping over many periods, matching the superlinear scaling of the Floquet state.

The stability of the approximation is studied beyond the high-frequency regime in Figs.~\ref{fig:Floquet_vs_coherent_vs_high_freq}(b),(c).  We compute the time-averaged work per period and per spin $\overline{W}_i/(NT)$ over 100 cycles for $\omega_1=\omega_2$ (b) and $\omega_2=\omega_2$ (c). Both high-frequency state and coherent state agree with the superlinear increase of the Floquet state for  $JN\lesssim \omega$. As $N$ grows larger, pumping of the high-frequency state then goes to zero $\overline{W}_i/(NT)\rightarrow 0$. Conversely, the work output of the  spin-coherent state has a qualitatively similar behavior as the Floquet state, settling to constant work per spin and per cycle (albeit lower than for the Floquet state). 

\begin{figure}[b!]    
    \centering
    \includegraphics[width=\linewidth]{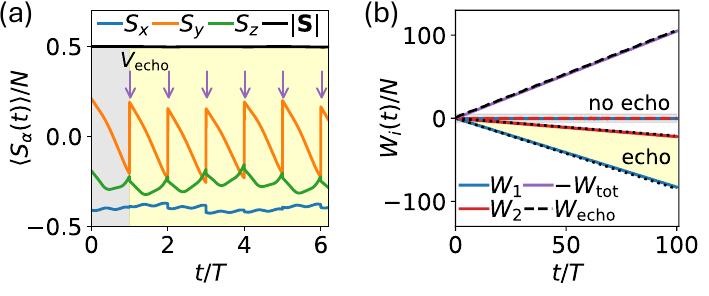}
    \caption{Stabilization of work states via echo protocol: An instantaneous pulse $V_\mathrm{echo}$ is applied after each cycle [purple arrows in (a)]. (a) Polarization dynamics. The system remains in a spin-coherent state ($|\mathbf{S}|\simeq 1$) within the period (gray shaded). The pulse then reverses the evolution. (b) Work output. Without echo, no work is produced (gray shaded). With echo (yellow shaded), work states are stabilized in the steady state (dotted black). The energy added by the echo (dashed black) is fully converted to the net work of the drives (purple).  Parameters: $N=16$, $J=0.25$, $\mathbf{h}_0=0$, $\mathbf{h}_1=\mathbf{\hat{y}}$, $\mathbf{h}_2=\mathbf{\hat{x}}$, $\omega_1=20\pi$,    $\omega_2=10\pi$, $\phi_1=0$, $\phi_2=\pi/2$, $\Delta \phi=-0.212\pi$, $\Delta  \theta=0.339\pi$.}
    \label{figure_echo}
\end{figure}

\subsection{Stabilizing work states by echo}
\label{sec:echo}

We now show that superlinear  pumping in the steady state can also be realized for work states. As previously noted, work states are not quasi-stationary, and thus require resetting after each driving period. Stabilization of work states is appealing because they also scale superlinearly, but generate substantially larger work compared to Floquet states. 

Here, we propose a periodic protocol that approximately resets them in a unitary manner [Fig.\ \ref{figure_echo}(a)]. We first evolve the system initialized in a work state over one period. At the end of the period, we apply a fixed, instantaneous pulse  
\begin{align}
    V_\mathrm{echo} = e^{i \Delta\phi S_z}e^{i \Delta\theta S_y},
    \label{eq:pulse}
\end{align}
which is parameterized by the polar and azimuthal angles $ \Delta\phi$ and $\Delta\theta$. The protocol is motivated by the polarization dynamics of the optimal work state,  $\ev{\mathbf{S}(t)}$ [Fig.~ \ref{figure_echo}(a), gray]. We observe that the total spin norm $\abs{\expval{\mathbf{S}(t)}}$ remains approximately maximal and conserved throughout the evolution over one cycle (gray area). This indicates that in the superlinear regime, time-evolved work states are close to spin-coherent states Eq.\ \eqref{eq:coherent spin states} with time-dependent angles $\phi(t)$ and  $\theta(t)$. The purpose of the pulse in Eq.\ \eqref{eq:pulse} is to time-reverse the one-period dynamics of the drives. This can be achieved by choosing the rotation angles $\Delta\phi=\phi(T)-\phi(0)$ and $ \Delta\theta=\theta(T)-\theta(0)$ in Eq.\ \eqref{eq:pulse}.

Computing the work of the individual drives, we find that the echo-induced stabilization of work states  generates the optimal work output in the steady state [Fig.\ \ref{figure_echo}(b)]. In contrast to Floquet states, the work performed by the individual drives does not need to sum to zero. The work supplied by the pulse after $N_c$ periods can be calculated as the instantaneous energy change
\begin{align}
    W_\mathrm{echo}= \sum^{N_c}_{n=1}\ev{\left[V_\mathrm{echo}^\dagger H(T)V_\mathrm{echo}-H(T)\right]}{\psi(nT)}.
\end{align}
The echo work $W_\mathrm{echo}$ exactly cancels the work by the two drives as shown in Fig.\ \ref{figure_echo}(b), restoring the energy balance  $W_1+W_2+W_\mathrm{echo}=0$.

\section{Conclusion}
\label{sec:conclusion}

In this work, we introduced a quantum charger that generates a steady-state energy flow between periodic drives. Realizing the charger as a spin chain, interactions induce collective charging  where energy pumping is superlinearly amplified with system size. Superlinear charging relies on cooperative motion of the constituent particles, converting the available interaction energy between the drives.

Remarkably, the effect holds for a wide range of realistic long-range interactions, realized, e.g., in trapped-ion experiments \cite{Britton2012,Islam2013,Zhang2017}.  Qualitatively, it is insensitive to the sign or nature of the interactions, as long as they are long-ranged. We find that nonlinear pumping persists up to a critical system size $N^\ast$, where the drive becomes strong and the charger splits into independent subsystems. The existence of $N^\ast$ suggests an efficient construction strategy. Beyond this size, one can partition a large charger into smaller sub-units and then scale up, rather than increasing the system size of a single unit. This construction requires only finite-box interactions with range $N^\ast$.  

The key quantity of our analysis is the work operator, whose largest eigenvalues set the maximum work output. Its eigenstates (work states) are superpositions of Floquet states and show superlinear charging, but pump significantly more than Floquet states.  Thus, while not required for superlinear charging, interference can significantly enhance the work. We demonstrated that work states can be coherently stabilized by echo pulses. 

In the steady state, the maximum performance is achieved in a Floquet eigenstate, and is given by the expectation value of the work operator in such a state. We find that classical spin-coherent states approximating the optimal Floquet states closely reproduce the quantum optimum in the collective pumping regime. Thus, the Floquet framework naturally allows for work optimization in the interacting case, which is a linear problem for a quantum spin chain but a nonlinear one in the classical case. 

Superlinear charging occurs in the high-frequency limit and is most efficient for drives with small, commensurate frequency ratios. It allows substantial energy transfer between drives but does not enable efficient frequency conversion. Extending superlinear charging to schemes capable of frequency conversion remains an open theoretical challenge, complementing recent experimental advances in quantum transducers \cite{Andrews2014,Zhao2025,Mirhosseini2020,Wang2022,Xie2025,Kumar2023}. 

Future work should also explore the resilience of the proposed charger against dissipation and disorder, as well as treating the drives as quantized cavity fields \cite{Nathan2019}. Although we focused on the LMG model, other many-body models may further optimize the charger. Further, it would be interesting to relate the superlinear charging effect to time crystals \cite{Wilczek2012,Khemani2016,Else2016,Sacha2018,Khemani2019,Else2020,Zaletel2023,Penner2025}. Time crystals are characterized by periodic many-body oscillations and can be realized in long-range settings \cite{Russomanno2017,Pizzi2021,Pizzi2019}. This suggests that they may serve as a useful resource for collective steady-state charging, in addition to existing practical applications in sensing
\cite{Montenegro2023,Pavlov2023,Cabot2024,Iemini2024,Yousefjani2025,Gribben2025,Tsypilnikov2025,Sahoo2025}.

\begin{acknowledgements}
HS thanks Molly Smith for discussions on trapped ions and gratefully acknowledges the hospitality of Caltech, where part of this work was conducted. Work at Freie Universit\"at Berlin was supported by Deutsche Forschungsgemeinschaft
within CRC 183 (project C03)
and the German Excellence Strategy - EXC3112/1 - 533767171 (Center for Chiral Electronics). YP is supported by the US National Science Foundation (NSF) Grants  No.\ PHY-2216774 and No.\ DMR-2406524. GR is grateful for support from the Simons Foundation, the Institute for Quantum
Information and Matter, an NSF Physics Frontiers Center (PHY-2317110), and the AFOSR MURI program, under agreement number FA9550-22-1-0339. This work was performed in part at Aspen Center for Physics, which is supported by National Science Foundation, grant PHY-2210452. 
\end{acknowledgements}

\section*{Data availability}
The data and code supporting this study are available on Zenodo~\cite{Schmid2026_zenodo}.

\appendix

\section{Further analysis on the scaling of $W$ for realistic interactions\label{app:fintie_interaction}}
To evaluate how good are our scaling estimate for the more realistic power-law interactions,
we plot the data similar to the ones in Fig.~\ref{fig:long_range} at a log scale. Due to the limitation of the numerical methods, we are only able to simulate up to 14 spins. We expect to get a better scaling behavior in a much larger system.

\section{High-frequency scaling of the work for $(p,q)=(1,2)$}
\label{app:high freq work}

We elaborate on the scaling of the work for $\omega_1/\omega_2=p/q=1/2$. In this case, the scaling collapse Eq.\ \eqref{eq:work freq scaling} predicts $W_{i,\alpha}(T)\propto \frac{N^2}{T^3}$. To recover this scaling from Eq.\ \eqref{eq:work Heff}, we identify terms in the high-frequency expansion that (a) contain a factor of $H_0$, yielding an $N^2$ scaling, and (b) exhibit dependence on $\phi_i$, which generically remains finite after taking the derivative $\partial_{\phi_i}$ (except in fine-tuned cases).

Up to order $\mathcal{O}(1/\omega^2)$, the work vanishes. The zeroth- and first-order terms satisfy only one of the above conditions. At second order [Eq.\ \eqref{eq:Heff}], the relevant commutators are of the form $[[H_p, H_0], H_{-p}]$ and $[[H_q, H_0], H_{-q}]$, but there are no mixed terms involving both $p$ and $q$. Consequently, no $\phi_i$-dependence arises at this order.

At third order, the high-frequency Hamiltonian of the van-Vleck expansion is \cite{Eckardt2015,Ikeda2021} 
\begin{widetext}
\begin{align}
H_{\mathrm{eff}}^{(3)} = 
&\sum_{m \neq 0} \frac{[[[H_{-m}, H_0], H_0], H_m]}{2 m^3 \omega^3} 
+ \sum_{m \neq 0} \sum_{\substack{n \neq 0 \\ n \neq m}} 
\frac{[[[H_{-m}, H_0], H_{m-n}], H_n]}{3 m^2 n \omega^3} 
+ \sum_{m \neq 0} \sum_{\substack{n \neq 0 \\ n \neq m}} 
\frac{[[[H_{-m}, H_{m-n}], H_0], H_n]}{4 m n^2 \omega^3} \notag \\
- &\sum_{m,n \neq 0} 
\frac{[[[H_{-m}, H_m], H_{-n}], H_n]}{12 m n^2 \omega^3} 
+ \sum_{m \neq 0} \sum_{\substack{n \neq 0 \\ n \neq m}} 
\frac{[[H_{-m}, H_0], [H_{m-n}, H_n]]}{12 m^2 n \omega^3} 
+ \sum_{m,n \neq 0} \sum_{\substack{l \neq 0 \\ l \neq m,n}} 
\frac{[[[H_{-m}, H_{m-l}], H_{l-n}], H_n]}{6 l m n \omega^3} \notag \\
+ & \sum_{m,n \neq 0} \sum_{\substack{l \neq 0 \\ l \neq m-n}} 
\frac{[[[H_{-m}, H_{m-n-l}], H_l], H_n]}{24 l m n \omega^3} 
+ \sum_{m,n \neq 0} \sum_{\substack{l \neq 0 \\ l \neq m,n}} 
\frac{[[H_{-m}, H_{m-l}], [H_{l-n}, H_n]]}{24 l m n \omega^3}.
\end{align}
\end{widetext}
The relevant commutator is $[[[H_{-q},H_0],H_p],H_p]\propto e^{i(\phi_2-2\phi_1)}$ which satisfies both conditions (a) and (b), and therefore yields a nonvanishing contribution to the work.

While explicit expressions for the van Vleck expansion beyond third order are, to our knowledge, not available, its structure up to third order reveals a clear pattern: in each term, the indices appearing in the nested commutators sum to zero. Physically, this corresponds to conserving the net number of absorbed and emitted “photons” from the drives, such that $l p = m q$ with integers $l,m$, where $l + m = n$ is the perturbative order. This observation provides a natural justification for the scaling collapse in Eq.\ \eqref{eq:work freq scaling}.

\begin{figure}[t!]
    \centering
    \includegraphics[width=0.45\textwidth]{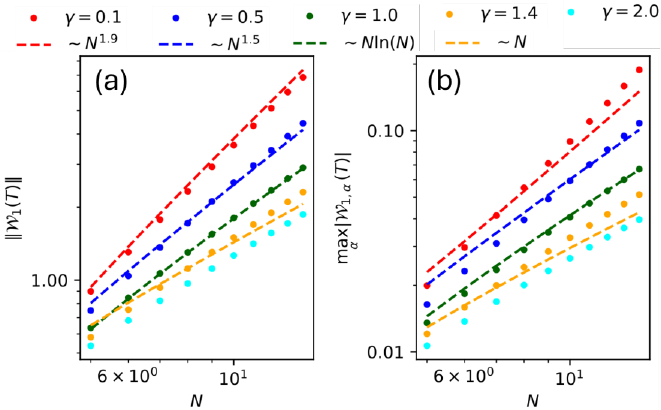}
    \caption{(a) Maximum work, and (b) work of optimal Floquet state, for drive 1 over a period as a function of system system size for different power-law interactions parametrized by $\gamma$, plotted at a log scale, with dashed guided lines for comparison. The parameters are the same as in Fig.~\ref{fig:long_range} (a) and (b).}
    \label{fig:powerlaw_scaling}
\end{figure}

\section{Low-frequency limit: semiclassical dynamics}
\label{app:Low-frequency limit: semiclassical dynamics}

We solve the equation of motion Eq.\ \eqref{eq:LLG} in the low-frequency limit $\omega \ll JN,h$. We also assume $h\ll JN$. We expand the spin in powers of $\omega$,
\begin{align}
\mathbf{s}(t)=\mathbf{s}_a(t)+\mathbf{s}_d(t) +\order{\omega^2},
\end{align}
where $\mathbf{s}_a(t)$ is the adiabatic solution and $\mathbf{s}_d(t)$ is the first diabatic correction. The adiabatic solution fulfills 
\begin{align}
    \mathbf{h}_{\mathrm{eff,a}}(t)\times
    \mathbf{s}_{a}=0.
\end{align}
with $\mathbf{h}_\mathrm{eff,a}(t)
 =   2Js_{z,a} \hat{\mathbf{z}}+\mathbf{h}(t)$. The effective adiabatic field $\mathbf{h}_\mathrm{eff,a}$ differs from $\mathbf{h}_{\mathrm{eff}}$
by $s_z$ replaced with $s_{z,a}$. Solving self-consistently with the ansatz $\mathbf{s}_{a}(t)= \frac{N}{2}\mathbf{\hat{h}}_{\mathrm{eff,a}}(t)$ yields 
\begin{align}
\mathbf{s}_{a}(t)\simeq \frac{1}{2J}
    \big(
        h_x\mathbf{\hat{x}}
        +
        h_y\mathbf{\hat{y}}
     +
     \left(JN+h_z\right)\mathbf{\hat{z}}\big).
\end{align}
Thus, the phase-derivative  $\partial_{\phi_i}s_z$ vanishes at leading order in $N$ and the nonlinear contribution to the work is zero.

Next, we show that also the first diabatic correction is phase-independent
\begin{align}
    \dot{\mathbf{s}}_a&=\mathbf{\hat{h}}_{\mathrm{eff,a}}\times \mathbf{s}_{d}
    -  \mathbf{s}_{a} \times (2Js_{d,z}\hat{\mathbf{z}}).
\end{align}
Plugging in $\mathbf{s}_a$ and using
$\dot{\hat{\mathbf{a}}}
    =
  \mathbf{\hat{a}}\times 
    (\dot{\hat{\mathbf{a}}}
     \times \mathbf{\hat{a}})$     
,
we find the first diabatic correction (written in components)
\begin{align}
    (\mathbf{s}_{d})_i(t)
    &=
    \frac{N}{2}
    \left(\dot{\hat{\mathbf{h}}}_\mathrm{eff,a}
    \times
    \mathbf{h}_\mathrm{eff,a}\right)_i
    \begin{cases}
        1\phantom{\frac{1}{1-NJ/h_\mathrm{eff,a}}}\qquad i=x,y,
        \\
        \left(1-\frac{NJ}{h_\mathrm{eff,a}}\right)^{-1}\phantom{1} \, i=z.
    \end{cases}
\end{align}
The spin lags behind the field in a direction both perpendicular to $\mathbf{h}_\mathrm{eff,a}$ and $\dot{\hat{\mathbf{h}}}_\mathrm{eff,a}$. To find the scaling, we note that in the transverse direction to the  exchange field, the effective adiabatic field has strength $\order{1}$. In contrast in the longitudinal direction, we have $h_\mathrm{eff,a}\sim \order{N}$. However, $s_{d,z}$ is suppressed by a factor $(1-NJ/h_\mathrm{eff,a})^{-1}\simeq -h_z/(NJ)$. As a result, also the first diabatic correction $\mathbf{s}_d$ is linear in $N$ in the low-frequency limit.

%
%

\end{document}